# RESULTS ON QUIESCENT AND POST-DISRUPTION RUNAWAY ELECTRONS MITIGATION EXPERIMENTS AT FRASCATI TOKAMAK UPGRADE

D. CARNEVALE[1], P. BURATTI[2], W. BIN[5], F. BOMBARDA[2], L. BONCAGNI[2], C. PAZ-SOLDAN[7], L. CALACCI[1], M. BARUZZO[2], M. CAPPELLI[2], C. CASTALDO[2], S. CECCUZZI[2], C. CENTIOLI[2], C. CIANFARANI[2], S. CODA[6], F. CORDELLA[2], O. D'ARCANGELO[2], J. DECKER[6], B. DUVAL[6], B. ESPOSITO[2], L. GABELLIERI[2], C. GALPERTI[6], S. GALEANI[1], S. GARAVAGLIA[5], G. GHILLARDI[2], G. GRANUCCI[5], M. LENHEN[4], D. LIUZZA[2], F. MARTINELLI[1], C. MAZZOTTA[2], F. NAPOLI[2], E. NARDON[3], F. OLIVA[1], L. PANACCIONE[2], M. PASSERI[1], C. POSSIERI[1], G. PUCELLA[2], G. RAMOGIDA[2], A. ROMANO[2], M. SASSANO[1], U. A. SHEIKH[6], O. TUDISCO[2],
the FTU team* and EUROfusion MST1 team[#]

[1] Dipartimento di Ing. Civile ed Informatica, Università di Roma "Tor Vergata", 00133, Rome, Italy
[2] ENEA, Fusion and Nuclear Safety Department, C. R. Frascati, Via E. Fermi 45, 00044 Frascati (Roma), Italy
[3] CEA, IRFM, F-13108, Saint Paul-lez-Durance, France
[4] ITER Organization, Route de Vinon sur Verdon, 13115 St Paul Lez Durance, France
[5] Istituto per la Scienza e Tecnologia dei Plasmi, CNR, via Cozzi 53, 20125 Milan, Italy
[6] Ecole Polytechnique Fédérale de Lausanne, Swiss Plasma Center, Lausanne, Switzerland
[7] Columbia Applied Physics and Applied Mathematics Department, 200 S.W. Mudd 4701 New York, 10027

* See the appendix of G. Pucella et al., Proc. 26th IAEA FEC, Kyoto, Japan, 2016
[#] See the author list of B. Labit et al. 2019 Nucl. Fusion 59, 0860020
Email: daniele.carnevale@uniroma2.it

**Abstract**

Results from the last FTU campaigns on the deuterium large (wrt FTU volume) pellet REs suppression capability, mainly due to the induced burst MHD activity expelling REs seed are presented for discharges with 0.5 MA and 5.3T. Clear indications of avalanche multiplication of REs following single pellet injection on 0.36 MA flat-top discharges is shown together with quantitative indications of dissipative effects in terms of critical electrical field increase due to fan-like instabilities. Analysis of large fan-like instabilities on post-disruption RE beams, that seem to be correlated with low electrical field and background density drops, reveal their strong RE energy suppression capability suggesting a new strategy for RE energy suppression controlling large fan instabilities. We demonstrate how such density drops can be induced using modulated ECRH power on post-disruption beams.

1. INTRODUCTION

Runaway electrons (RE) are one of the major concerns for ITER operations. In tokamak devices with high fusion rates and high current, a loss of plasma confinement can lead to runaway electron formation (possibly up to 12 MA in ITER) via primary and secondary RE generation mechanisms [1,2]. Runaway electrons that form after a disruption can attain high energies, mainly induced by the large electrical field at current quench (CQ), and the small pitch angle, that may then deposit unsustainable power fluxes upon the plasma facing components may result in deep melting within the tokamak structure. The main strategies for RE mitigation rely on increasing collisionality to avoid RE beam formation or to quickly dissipate the electron energy where RE are formed [1,8]. RE current dissipation is unavoidably associated with undesirable fast-growing vertical displacement events, i.e. fast RE energy dissipation by heavy-Z material injection is often accompanied by fast current decay leading to uncontrollable VDEs: the result is a competition among vertical displacement, energy dissipation and electromechanical loads that may not have ITER feasible solutions. Research is seeking further techniques that may be used in combination with MGI/SPI, such as a dedicated control strategy [3,4], 3D stochastic fields by resonant magnetic perturbation [7] and generating other instabilities [9]. Recent results obtained at DIII-D, and further investigated at JET, show that large RE currents, driven by the central solenoid following deuterium injection (SPI), induce current-driven (low safety factor) kink instabilities with extremely fast RE beam losses with no sign of harmful energy deposition, opening a path for an alternative RE mitigation strategy [3]. Continuing the studies of past years at FTU [4,5] we explored a number of alternative solutions for RE mitigation.

## 2. PELLET INJECTIONS ON QUIESCENT SCENARIOS

FTU has performed a large number of pulses with significant RE seeding population named *quiescent RE scenarios*. Such scenarios were achieved by a low $D_2$ gas prefill and reducing the electron density $n_e$ reference below 3E19 m$^{-3}$. Here the electric field usually overcomes that for runaway generation [5] by Dreicer mechanism and runaway electrons forms with a rate that is usually proportional to the electric field i.e., proportional to the flux loop voltage generated by the active coils (central solenoid and V coil). In such non-disruptive scenarios, the growth rate that is inversely proportional to electron density $n_e$, lead to complete RE suppression. There are other factors such as radiative losses by synchrotron emissions [6] that modify the RE number and energy growth rate as well as confinement losses due to error fields, drift and instabilities. FTU's $D_2$ pellet injector on the equatorial plane and capable of injecting up to 4 pellets of different size: two $1 \times 10^{20}$ $D_2$ atoms at about 1200 m/s and two $2 \times 10^{20}$ $D_2$ atoms at about 1000 m/s that require about 0.3 ms to reach the plasma core. The main use of such system was to raise the density (fueling) up to $8 \times 10^{20}$ with a maximal current $I_p$ up to 1.2MA (8T) [14]. While pellets are able to induce disruptions with RE beam generation, named appropriately "killer" pellets [15], when launched on fully formed RE beam, they can reduce drag collisionality and consequently the current dissipation rate [16] [17] improving RE beam controllability, thus suggesting a possible strategy for RE beam controlled shutdown or, more exotically induce a sudden kink instability that results in uniform energy deposition [3] with no sign of high temperature spots so often associated to RE beam impacts on the vessel.

### 2.1. Runaway electron growth after pellet injections

Herein, we discuss two different effects induced by pellet injections on quiescent scenarios. When a single pellet is injected into a flat-top discharge with a considerable RE seed population, if the RE population is not lost during burst MHD activity, we observed an increase of RE seed up to the saturation of the NE213 diagnostic that is sensitive to gamma-rays and neutron ($\gamma + n$) induced by in-flight or lost REs [18]: Cherenkov probes lead us to conclude that RE loss during this growth phase are not larger than losses before the pellet injection. The signal ($\gamma + n$) in the logarithmic plot of Fig. 1 should mirror the growth of RE seeding. For the two discharges #43654 and #43651 a single 1E20 deuterium pellet was injected at 0.3 s (the pellet effects are visible within 6ms from the triggering signal) whereas for the shot #43539 two pellets, of different size (1E20 + 2E20), were launched, almost simultaneously at 0.7 s, leading to a higher density increase. After the pellet cools the plasma, the current control loop reacts to the plasma current $I_p$ that smoothly decreases, due the increased resistivity, by augmenting the current rate on the ohmic transformer, and consequently, the electrical field ($V_{loop}$). This electrical field increases further REs while the density decrease give rise to MHD activity that, about 60-70ms after the density peak, leads to a disruption with formation of a RE beam. This phenomenology has been exploited on FTU to reliably produce post-disruption RE beams without MGI.

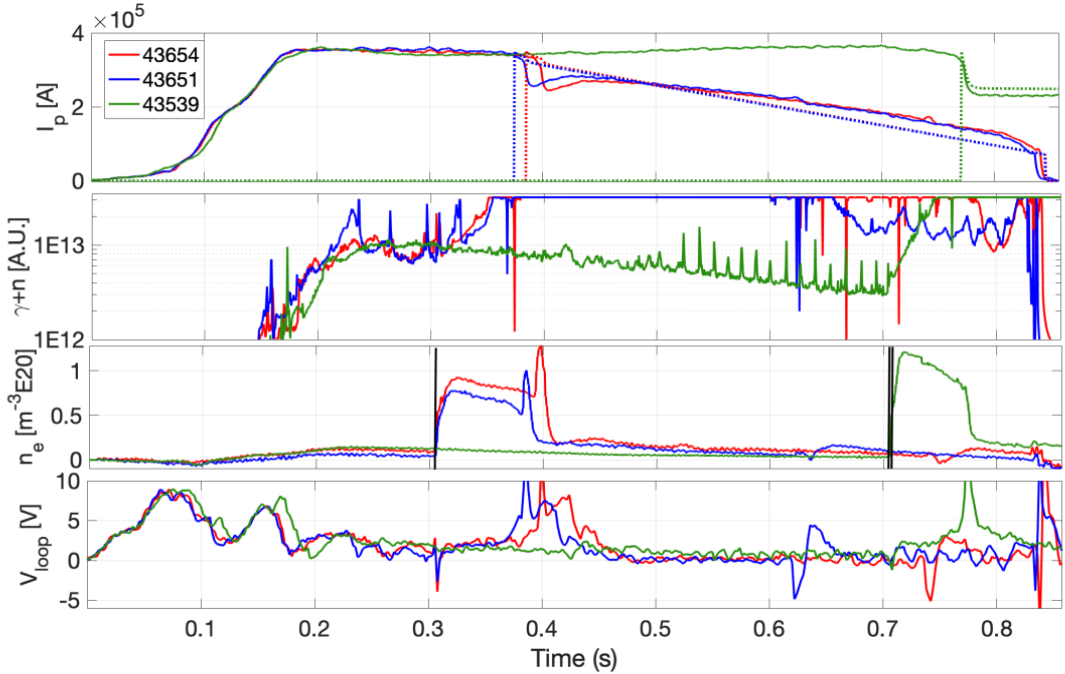

*Fig. 1. Pellet injections on quiescent REs leading to post-disruption RE beam formation.*

This recipe to generate RE beams also provides qualitative information on RE dynamics prior to disruption. Indeed, it is possible to employ the difference of signals NE213 and BF3, the last of which counts neutrons, to define $\bar{\gamma} = NE213 - BF3 \approx \gamma$ that may be considered as rough estimate of the number/energy of the in-flight RE seed. The critical electrical field $E_r$ below which no RE could form [5] as

$$E_r = \frac{n_e e^3 ln\Delta}{4\pi\varepsilon_0^2 m_e c^2}, \qquad (1)$$

that is at least 5 (5-20) times lower than the normalized electrical field $\bar{V}_{loop} = V_{loop}/2\pi R_0$, with $R_0 = 0.96$ m, above which REs are seen to form in FTU i.e., $\bar{\gamma} > 0$ (possibly increasing in time) when $\bar{V}_{loop} \geq \sigma E_r$ and $\sigma \geq 5$ in FTU ($B_T \in [4T, 8.1T]$). In the next we show that avalanching need to be invoked to fit the growth of $\bar{\gamma}$ in post-pellet RE quiescent scenarios with $B_T$ =5.3T and $I_p$=0.35MA.

Neglecting synchrotron radiation and deconfinement losses [5,6,19] with unrealistically low $\sigma$=1, we should be able to provide an upper bound of the $\bar{\gamma}$ growth relying only on primary generation mechanism i.e., the estimate $\hat{\gamma}_{Dreicer,\sigma} = \int \theta(\bar{V}_{loop} - \sigma E_r)$, for an opportune scaling parameter $\theta > 0$, should provide un upper-bound of $\bar{\gamma}$ with $\sigma$=1. We set the parameter $\theta$ = 1E15 in order to fit the initial growth of $\bar{\gamma}$ as shown in the time traces (black dashed) $\hat{\gamma}_{Dreicer,1}$ ($\sigma$=1) of Fig. 2, selections $\sigma > 4$ yield negative slopes in some such post-pellet quiescent RE discharges, as for #43539 where the $\hat{\gamma}_{Dreicer,4}$ rate (black dashed curves in Fig. 2) is almost zero. Nevertheless, even with such low selection $\sigma = 1$, the estimate $\hat{\gamma}_{Dreicer,1}$ runs below $\bar{\gamma}$ about 5 milliseconds before the signal NE213 saturates.

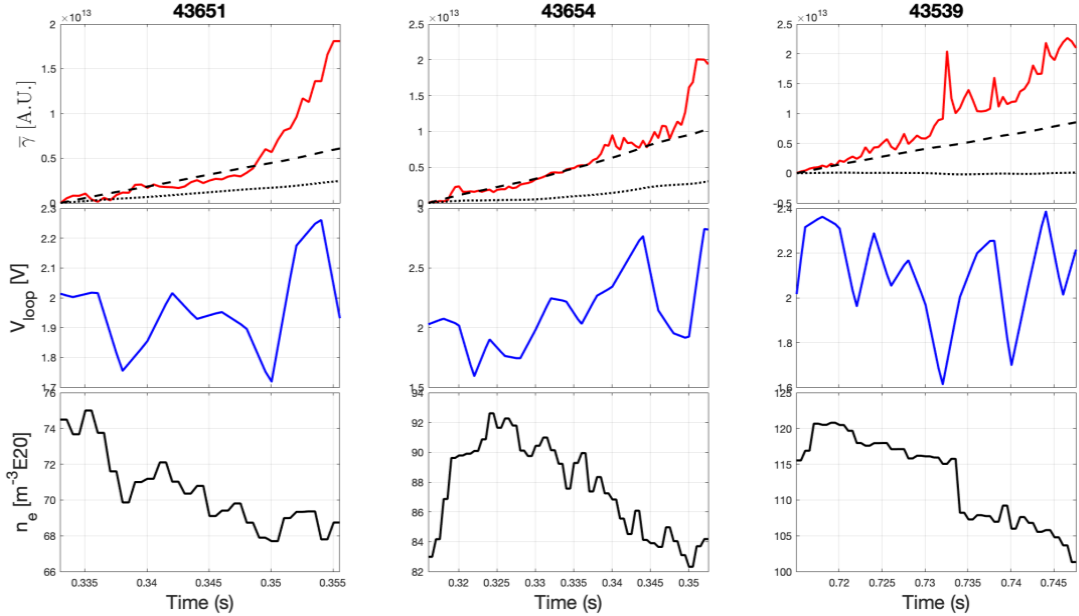

*Fig. 2. In the first row are shown the experimental $\bar{\gamma}$ (red) compared with $\hat{\gamma}_{Dreicer,4}$ (black dots) and $\hat{\gamma}_{Dreicer,1}$ (black dashed) with $\theta$ = 1E15.*

To observe RE growth in FTU, it is often necessary to maintain $\bar{V}_{loop} \geq \sigma E_r$, with $\sigma \geq 5$. During the time intervals following pellet injection, if we consider $\hat{\gamma}_{Dreicer,\sigma}$ with $\sigma \geq 5$ we would conclude that REs are suppressed. This indicates that other RE generation mechanisms, such as avalanche multiplication, are required to explain such an increase in $\bar{\gamma}$. In [19], the avalanche mechanism for flat-top current discharges was taken into account for FTU, but the estimated avalanche time-constant, from including runaway losses via radial diffusion, led to discard such an option. At least, in this specific "post pellet" phase, it seems that **avalanche multiplication is now necessary to explain the (superlinear/exponential) growth of $\bar{\gamma}$**. We cannot, a priori, exclude possible increments due to MHD activity, although this usually enhances REs losses. We do not include hot-tail RE generation since the time intervals for which analysis is performed are relatively far from the cooling effect of the pellet, although an initial increase of RE seeding could be provided by hot-tail generation mechanism during the pellet ablation. Note that in [29] the Authors describe a second electric field threshold that also appears to generate avalanche growth whereas herein we are below such a threshold.

## 2.2. Runaway electron growth in presence of fan instabilities

Another phenomenon arises when Dreicer generation mechanism is used to estimate the RE growth in presence of fan instabilities, which also increase their expulsion and decrease their power by enhanced synchrotron emission due to pitch angle increase [20, 21, 22].

In Fig. 3 we try to fit $\bar{\gamma}$ using $\hat{\gamma}_{Dreicer,\sigma}$ succeeding only with $\sigma = 33$ ($\theta = 1E15$) i.e., more than 2 times the $\sigma$ found in discharges with negligible fan instabilities, where $\sigma \in [5,15]$. In the shot #43539 of Fig. 3 low amplitude and high-frequency fan instabilities take place at early stage (about 0.25 s), expelling REs (red Cherenkov probe trace at the bottom of Fig. 3) and increase energy losses of seeding runaways by increasing their pitch angle and, consequently, their synchrotron radiation losses. After 0.45 s, when density decreases below 6E18 m$^{-3}$ (minimum of 2.7E18 before the pellet at 0.706 s) and the mean loop voltage decreases below 0.9 V, a low frequency, but high amplitude, fan instability is clearly seen in the NE213 signal (middle plot of Fig. 3) and the Cherenkov probes. When such large amplitude instabilities occur, even a value $\sigma = 33$ does not allow to capture the decrease of $\bar{\gamma}$ (energy losses and REs expulsion). At the end of the time interval of Fig. 3, as already discussed in Section 2.1, the estimated $\hat{\gamma}_{Dreicer,\sigma}$ would lead to fast RE suppression with $\sigma = \{15, 33\}$. This analysis shows that **fan instabilities strongly reduce the energy/number of runaway electrons** and may quantitatively be accounted for by at least **a doubling of the critical (experimental) electrical field $E_r$**. The fitting of an improved model that accounts for synchrotron radiation, avalanche multiplication and losses yield by instabilities is planned. These findings corroborate the mitigation effects of instabilities on REs such as Whistler waves reported in [31]

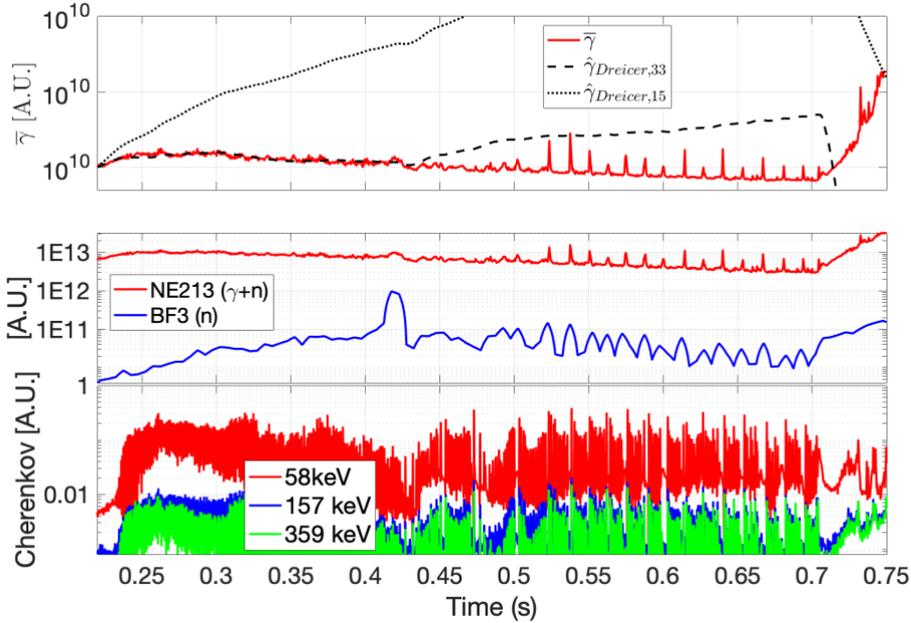

*Fig. 3. The experimental $\bar{\gamma}$ (red - top plot) for the shot #43539 compared with $\hat{\gamma}_{Dreicer,15}$ (black dots – top plot) and $\hat{\gamma}_{Dreicer,33}$ (black dashed - top plot) with $\theta = 1E15$. In the bottom plot the Cherenkov emission sensed by three probes colleting energies above 58 keV (red), 157 keV (blue), and 359 keV (green) produced by electrons leaving the plasma: small amplitude and high-frequency fan instabilities set in from the beginning (0.25 s) and sensibly increase in amplitude and decrease in frequency after 0.45 s, the time at which $\hat{\gamma}_{Dreicer,33}$ clearly diverges from $\bar{\gamma}$.*

## 2.3. Runaway electron expulsion by pellet injections

We analyze now the fast interactions between pellets and RE seed i.e., the loss of RE seed that are registered after pellets injection. In FTU, Laser Blow Off (LBO) injections using high-Z materials, especially Fe, have shown that LBO quite efficiently triggers bursts of MHD activity expelling REs [4] both in RE quiescent scenarios and for post-disruption RE beams when enough energy is transferred from the beam to the background plasma so it reaches the temperature necessary for particles ionization. The analysis on RE interaction with pellets on 0.36MA discharges requires deeper analysis as there are cases where REs are expelled (not mitigated by increased drag collisionality) and others that lead to disruptions, generating RE beams, as depicted in Fig. 5 ($B_T$ =5.3T). The picture is clearer on 0.5MA discharges (15 pulses): **a single or even multiple pellets injections into quiescent REs induced (complete) loss of REs without disruptions nor RE beams formation**. RE expulsion by bursts of MHD activity is within [8,100] ms of the first injection (multiple injections are performed within 50 ms following the first pellet). An example of such cleansing effect is shown in Fig. 4.

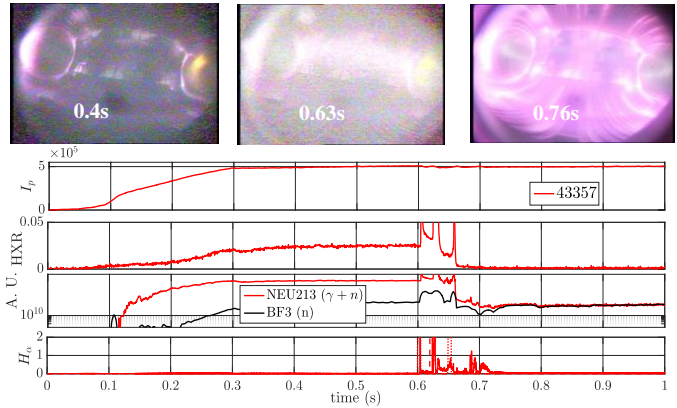
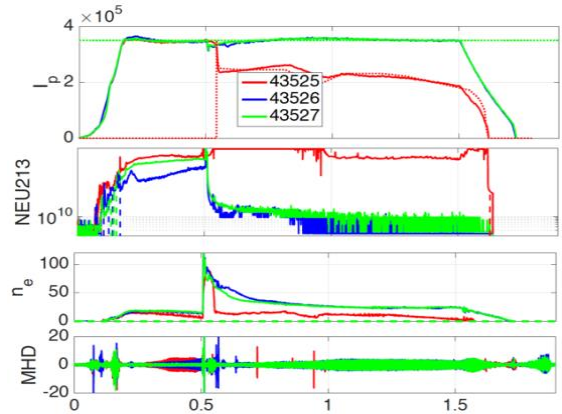

**Fig. 4:** RE mitigation on quiescent scenario injecting a small $D_2$ pellets 1E20@0.6s followed by 2E20@0.62s: REs are completely expelled. No post-disruption RE beams are generated by pellet injections into 0.5 MA discharges.

**Fig. 5:** Single 2E20 $D_2$ pellets are injected at 0.5s: complete RE loss for #43526 and #43527, post disruption RE beam for #43525.

In order to assess the effect of pellet injections in presence of loop voltages higher than that obtained during flat-top discharges i.e., to probe the effect of pellets on RE seed population with high loop voltage mimicking "slow" current quenches expected on ITER. The current was ramped from 360 kA to 500kA whilst simultaneously injecting single or multiple pellets yielding complete loss of REs followed by disruption. A possible explanation is that a higher loop voltage enhances MHD activity expelling REs and inducing a major disruption. **This is somewhat encouraging and suggests to repeating such experiments on other tokamaks for possible ITER predictions**. An example of such experiments is provided in Fig. 6 where the loop voltage is 2.8 V before pellet injection. Further analysis will be performed on pellet ablations for both flat-top and ramp-up scenarios exploiting fast H-alpha and diamond detectors [30].

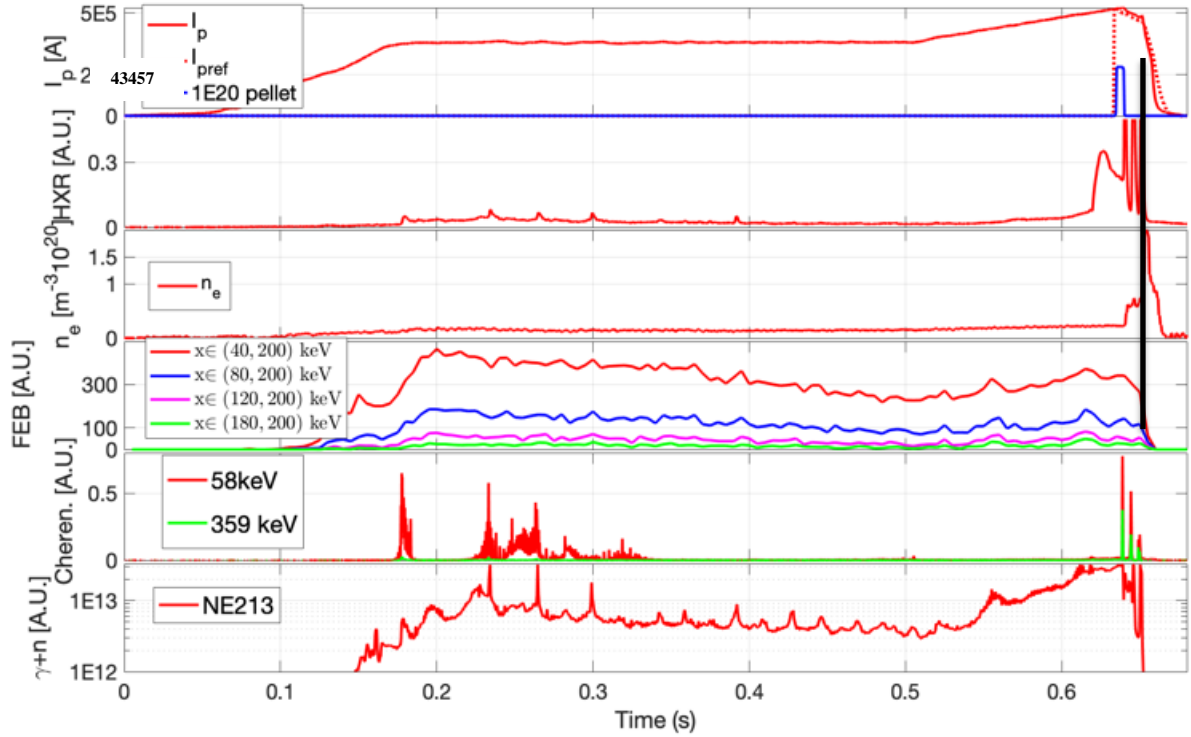

*Fig. 6. (Left - #43457) An example of pellet injection during current ramp-up discharges when the electrical field is above 2.5V: in all discharges burst MHD activity led to disruptions with complete RE loss. (Right - #42645) Iron LBO injected at 1.24 s generating as well burst MHD leading to (major) loss of REs.*

3. LARGE FAN INSTABILITIES ON POST-DISRUPTION RE BEAMS

Large fan instabilities, are now shown to strongly reduce the RE beam energy, even for post-disruption runaway beams and seem to provide a strong mechanism that swiftly decreases the RE beam energy and increases the background plasma temperature. Post-disruption RE beams often show sudden inward movements synchronous with negative spikes of the sensed loop voltage and intense spikes on HXR/Cherenkov detectors. Following this, the number/energy of the REs beam appears to suddenly decrease while the background plasma temperature increases. Such rapid inner movements registered by the magnetic coils and reconstructed using ODIN algorithm were, indeed, found to result from large fan instabilities. Fig. 7 shows shot #42747 in the right column where one notes a large inward movement with a large decrease of the reconstructed external beam radius ($R_{ext}$) at around 0.5 s, with a large negative spike in the loop voltage. A FEB camera, that records bremsstrahlung emissions of inflight suprathermal electrons with energy in the interval between 40 keV and 200 keV (e.g. Fig. 6) showed that the a suprathermal electron population disappeared immediately after the CQ indicating that only higher energy electrons, invisible to FEB, were present after the CQ. It is also important to note that, after this large fan event, the HXR signal decreases below its saturation. A second, milder, fan instability takes place at around 0.6 s. After the second event, the background plasma temperature increases up to values able to ionizing pellet and high-Z materials (LBO injections).

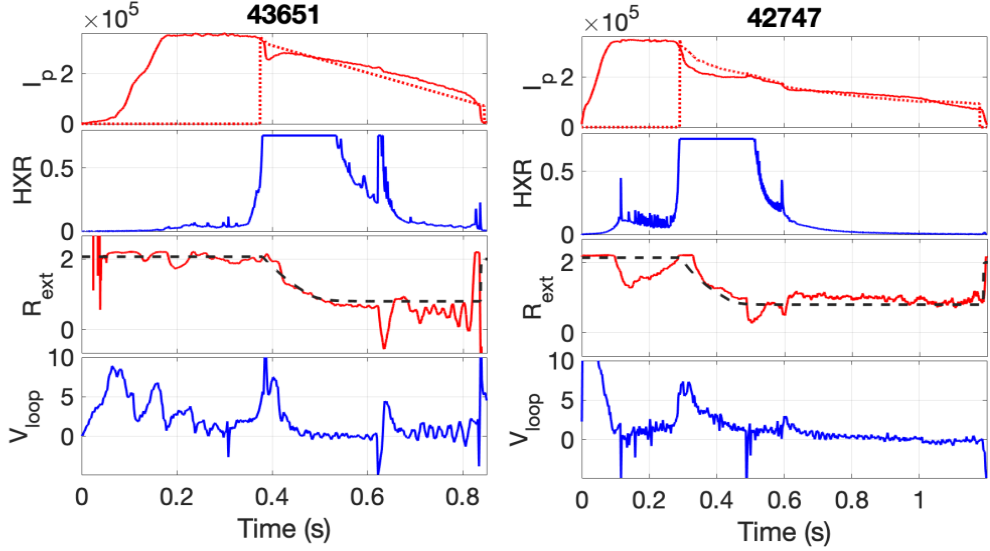

*Fig. 7. Time traces of pulses with clear signs of large fan instabilities associated to large inward beam movements.*

The time traces of the pulse #43651 in the left column of Fig. 7 show similar features: at about 0.621s a large RE beam inward movement is detected with corresponding HXR and loop voltage spikes. The HXR spike results from energetic RE interactions with the vessel, whereas the Vloop spike is mainly induced by flux changes of the RE beam that shrinks towards the high field side where the sensor coil is wounded. Final proof of this picture is provided by the images shown in Fig. 8. Before the large fan instability event at 0.621s in pulse #43651, the magnetic surfaces are close to the low field side. Synchrotron light sensed by the camera is well localized and the flux function shows a maximum towards the high field side (flux profiles are depicted in the top-right box of the Fig. 8 for different times - %e.eqlpsivr(time)). The electron density also displays similar profiles (bottom-right of Fig. 8 - %e.sidlvr(time)). Following the large fan instability, the RE beam rapidly moves towards the high field side as seen by magnetic surfaces and flux profiles. To stress that this sudden inward movement is associated with the fan instability, we recall that the radial outward shift of RE orbits depends on their mean kinetic parallel energy (momentum) as roughly approximated in [26] by $\Delta R_{RE} \approx \frac{\gamma m v_{\parallel} c}{eRB_p} = \bar{q} E_{RE,\parallel}/ecB_0$ (see [17, 28] for further considerations), where $E_{RE,\parallel}$ is the (mean) parallel kinetic energy of the runaway population, $\bar{q}$ the mean safety factor and $B_0$ the (mean) toroidal field. As $\bar{q}$ increases following the fan-like instability, the only explanation of this radially inward movement is that $E_{RE,\parallel}$ suddenly decreases, which is indeed the usual effect of fan-like instabilities. This inward movement it is not the same as the one at current quench where the control system is not able to reduce the vertical field (derivative constraints on coil voltage amplifiers) according to the current step.

Several publications [13, 24] note that the internal magnetic energy estimated from $E_{RE,mag} \approx \mu_0 l_i R_{RE} I_{RE}^2/4$, decreases due to an internal inductance decrease associated with the increased REs pitch angle after the fan-like event, which explains the large negative spike in the $V_{loop}$ signal. Some of the magnetic energy is transferred to background plasma that heats: this is an indirect observation since deuterium, as well as heavy-Z particles released by pellets and LBO injections, are ionized after such instabilities (see Fig. 9). Such ionization is not observed before fan-like instabilities and $D_2$ pellets are able to further reduce the background plasma temperature down to recombination levels.

To conclude, we have seen that such (large) anomalous doppler instabilities reduce the REs energy by: increased radiation losses due to increased pitch angle widens synchrotron emission annular spot size (bottom picture in Fig. 8), REs direct losses to the vessel detected by synchronous spikes of the Cherenkov/HXR probes, and energy exchange with the background plasma.

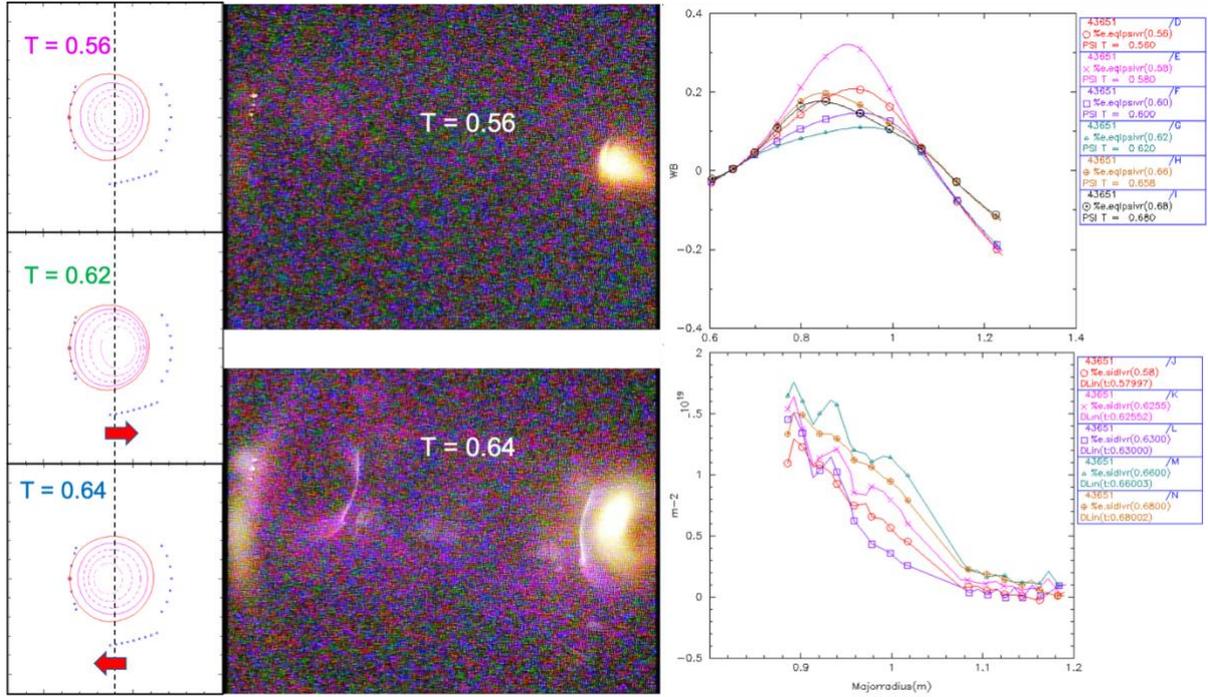

*Fig. 8. On the left the flux surfaces reconstructed by ODIN at different times, in the middle the camera picture of port 3 clearly showing the enlarged pitch angle after the fan-like instability at 0.621 s. On the top right are shown the normalized flux profiles vs the major radius and the electron density profile at the bottom right (shot #43651).*

The sensitivity of our diagnostics depends upon the amplitude of the anomalous doppler instabilities. For example, smaller and high frequency perturbations, usually found within quiescent plasmas with densities above 1E19 m$^{-3}$ or following large perturbations on post-disruption RE beams, are characterized by a sudden increase of ECE intensity, whereas the larger ones, during the early phase of post-disruption RE beam, are visible on the MHD Mirnov coils (since they induce sensible changes on the REs trajectories), as spikes on Cherenkov probes and small negative spikes on the loop voltage signal (see Fig. 9). The largest perturbations are also accompanied by a sudden radial inward movement and, ~interestingly~, occur certain time after the current quench onset: we infer that is the time required to slow down a large part of RE population until its momentum distribution is largely affected by the anomalous doppler instability. The time between the CQ onset and large fan-like instabilities, found to be in the range [70 ms, 270 ms], should depend on the loop voltage at CQ accelerating REs, density, drag collisionality by impurities, spatial losses i.e., it is the time interval required to shape the REs momentum distribution function to show the Parail-Pogutse instability involving a large part of runaway electron population. As reported in [24], no anomalous doppler instabilities affected REs above 4MeV in TFTR. Similarly, in FTU the fission chamber (FC) signal sensing photo-fissions induced by REs with energy higher than 6MeV impacting the wall, shown at the bottom of Fig. 9, goes to zero before such large instabilities are seen. Then, following the sequence described above, large fan instabilities seem to be triggered earlier when the sum of the central solenoid flux provided to the RE beam and the induced electrical field at CQ are smaller i.e., when the beam is less sustained by the central transformer or less accelerated during the CQ.

The HXR level (cfr. Fig. 7 or Fig. 9) often drops below saturation level after such instabilities: this drop should be instantaneous but the HXR diagnostic is affected by an exponential decay with time constant approximatively equal to 0.0917 s$^{-1}$. The HXR signal remains well below its saturation threshold and then goes to zero, without an increase at final loss, indicating that the REs energy has been dissipated during the current ramp-down.

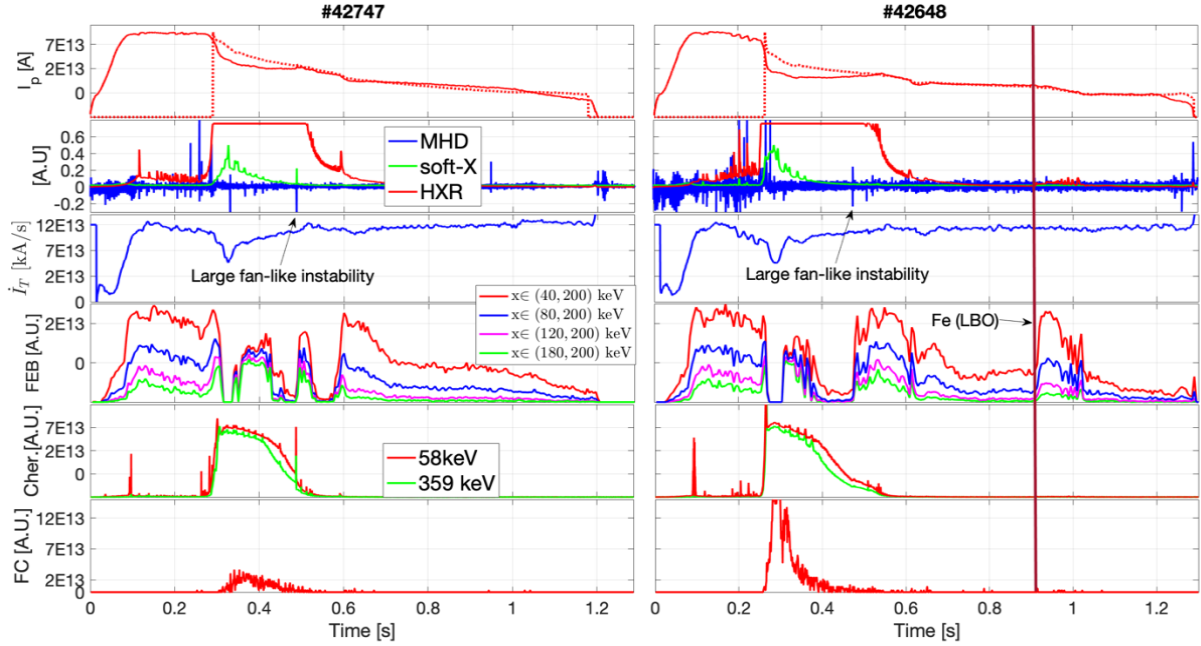

*Fig. 9. Post-disruption RE beams: large fan instability seem to largely dissipate the RE beam energy. The current derivative $\dot{I}_T$, proportional to the flux provided by the central transformer to sustain (negative values) the beam, is progressively reduced. In the shot #42648 an LBO injection with iron at 0.9 s, on a sufficiently hot background plasma, induces pitch angle scattering again by fan-like instabilities and current loss: note that sopra-thermal population (FEB) can be re-generated by the induced electrical field increase.*

The central horizontal line of the FEB camera (Fig. 9), senses different ranges of HXR produced by suprathermal electrons, revealing the charging phase before the large anomalous doppler: the pitch angle decreases and so the HXR sensed by the central line (also other view lines) of FEB and then the kinetic instability take place increasing the RE pitch angles. Accordingly, the Cherenkov probes feature small spikes superimposed on their traces.
On the shot #42648 LBO was injected into what remained of a post-disruption RE beam demonstrating that the background plasma has sufficient temperature to ionize the iron particles increasing electrons pitch angle by high-frequency anomalous doppler causing non-negligible current losses (expelled REs have energies below the sensing range of Cherenkov probes).
The 15 radial and 15 vertical FEB view lines can be used to estimate the beam movements and pitch angle modifications, even during the CQ: the suprathermal population with energies lower than 200 keV decreases as the electrical field quickly accelerates fast electrons to much higher energies (invisible to FEB) and the only view lines retaining large signals are those pointing at the higher vertical locations of the chamber (runaway vertical drift [26]). FEB signals also decrease when the RE beam pitch angle shrinks before the large fan-like instability.
During the CQ, the FC diagnostic, detects impacts of high energy (> 6MeV) REs on the vessel with no signs of vessel hot spots implying that during RE beam creation losses are not localized. The FC signal is much higher in the discharge #42648 due to a larger CQ inducing a higher accelerating electrical field.

From the previous analysis on the energy dissipative effects of large fan-like instabilities, we now propose that a complete RE to plasma conversion was achieved in TCV pulses #55069 and #55070, as initially introduced in [4], since the same indications of large fan-like instabilities described for FTU have been noted in these TCV discharges: after a controlled RE beam ramp-down, where OH coils current oscillations have been applied appositely to provide oscillating negative electric fields with the aim of breaking REs velocities below the critical (run-away) velocity by exploiting hysteretic effects [4]. As shown in Fig. 10, a sudden inward radial movement of the beam, associated with large negative electrical field and HXR spikes (as in FTU) are seen: REs completely disappeared (no HXR signal) in what no resembles a re-established ohmic plasma that survives for a further 50 ms (with a loss of about 15% of the current before the perturbation). This proposition is inspired as such large instability, after the controlled ramp-down, should have considerably decreased the REs energy below the critical point where electrons become run-aways (critical velocity), inducing their thermalization. To of Author's knowledge, although on FTU the RE beam energy is almost completely dissipated before the final loss, these TCV discharges are unique as no more signs of REs are present at the final loss that is, as of today, unique on a tokamak.

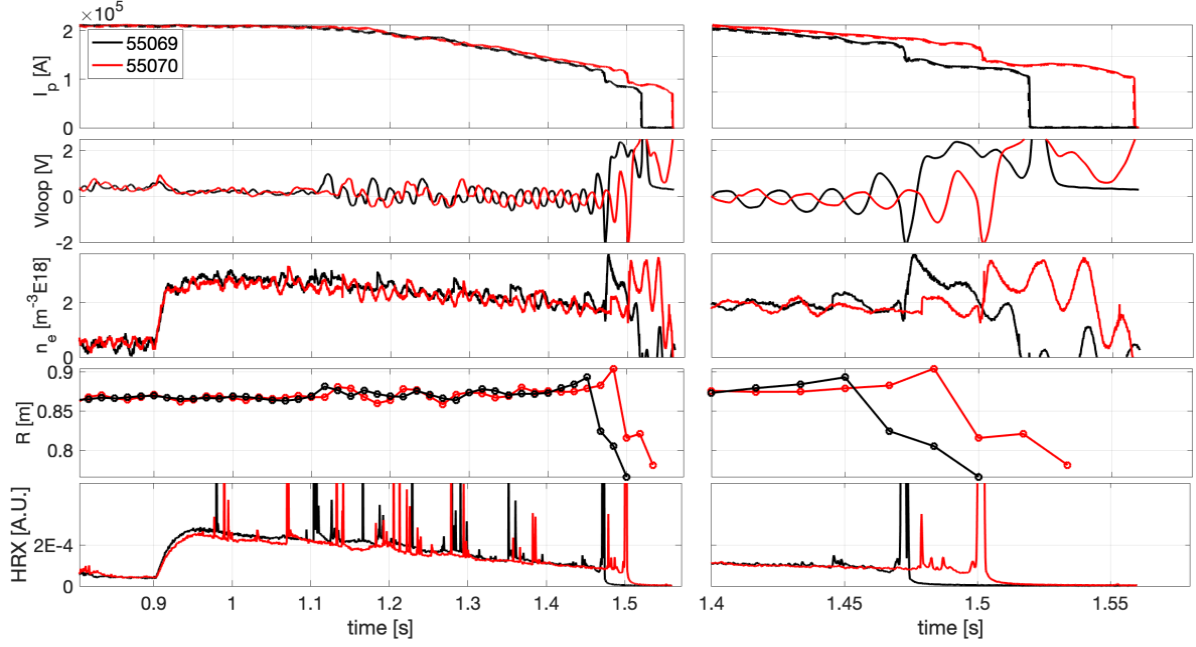

*Fig. 10. The two TCV discharges showing the same phenomenology of large fan-like instability in FTU and reaching complete disappearance of REs before the final loss.*

4. ECRH ON POST-DISRUPTION RE BEAMS

Few experiments have been conducted using ECRH on post-disruption RE beams in FTU. They do, however, reveal interesting properties that open the possibility of pacing large fan-like instabilities i.e., to actively control such instabilities and enhance RE energy dissipation. Fig. 11 shows a discharge that naturally disrupted at the start of the current flat top (360kA) and into which a large pellet (2E20) was injected during a slow CQ together with prompt current reference ramp-down triggered by HXR safety threshold (0.2, red time trace in the second plot from above). We are performing statistical analysis on a large set of discharges to assess whether such prompt reference current decrease (leading to electrical field decrease) in combination with a large pellet injection may result in what we can define as "weak" RE beam preventing the NE213 saturation generating one of the shortest time intervals (71 ms) of saturated HXR on post-disruption RE beam. The 4E20 $D_2$ injected pellet, whose penetration into the beam is tracked by H-alpha detectors, have slightly raised $n_e$ and could participate to avoid the formation of an energetic RE beam: there indications also in other discharges that pellet injected into CQ (very close) might lead to smaller current drop and then reduced electrical fields. At around 0.4 s, a first large anomalous doppler instability occurs inducing ionization of the deuterium that is being continuously injected by a gas valve (blue dashed line, top plot). Then, a series of "natural" and regular fan-like events with a period of 32 ms (31 Hz) appear to be trigger when density decreases below certain level, with a modified period that is forced by the action of injected power of ECRH, itself modulated at 50Hz. This phenomenon may be the indication that anomalous doppler instabilities can be triggered by decreasing densities down to a given value that should depend on $B_T$ and background temperature. The confirmation of such findings opens the path to another mitigation strategy that, if applied in the early RE beam phase, could allow faster REs energy dissipation to be used on a formed/forming RE beam and that could be feasible on large machines too such as ITER.

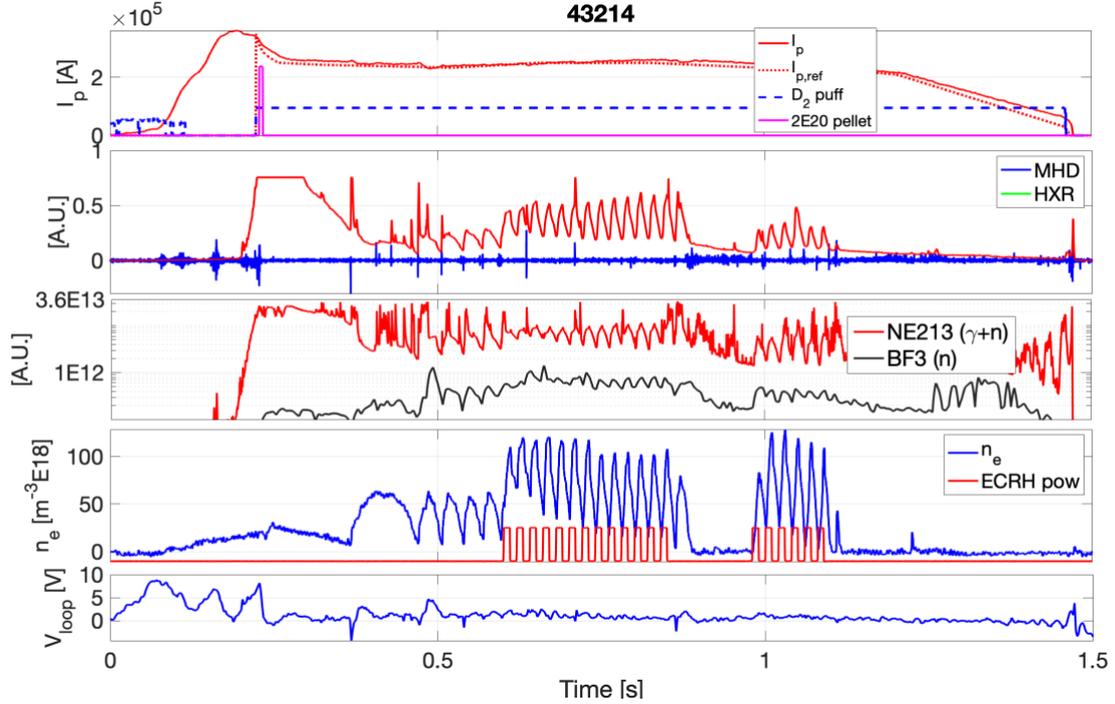

*Fig. 11. The discharge that might reveal the possibility of triggering large fan-like instabilities generating background decreasing density via modulated ECRH.*

5. CONLUSIONS

We have presented and discussed results obtained from FTU dealing with a number of aspects of RE beam modelling and mitigation strategies. We have shown that avalanches should be accounted for in RE dynamics even for quiescent discharges and that anomalous doppler instabilities play an important role in dissipating REs and re-distributing their energy. We reported the effectiveness of REs removal by (single/multiple) pellet injections on RE quiescent discharges with currents higher than 360 kA and REs complete loss accompanied by disruptions that do not lead to RE beams if pellets are injected during a current ramp. The latter results are obtained with an experimental setup to test the effect of pellet injections with loop voltage higher than in flat-top scenarios. Further studies will assess the pellet ablation for extrapolation to ITER. The important role of energy reduction and dissipation by large fan-like instabilities on post-disruption RE beams has been discussed and enough understanding to provide possible explanation for observations on the unique TCV discharges. ECRH effects on a post-disruption RE beam with the possibility of triggering large anomalous doppler instabilities controlling density ramp-down with ECRH is introduced.

The results from FTU experiments would suggest several avenues to palliate the RE threat. Prompt injection of large deuterium pellets, when a disruption is detected, can be combined to current reference ramp-down to limit energy transfer from flux loops to REs and possibly use ECRH to augment deuterium ionization leading density decrease that triggers large fan-like instabilities, strongly enhancing REs dissipation. These strategies can be combined during the longer current quenches such as those predicted for ITER in order to avoid RE beam formation. In case a RE beam forms after the CQ, then dedicated RE control tools have been shown able to gently ramp-down RE current acting on the central transformer and the aforementioned new strategies may also be considered to limit REs energy before the final control loss, estimated to be around 2MA for ITER.

ACKNOWLEDGEMENTS

We would like to thank the ITER organizations, and in particular Michael Lenhen, that supported with an important motivation letter the last experimental campaign of FTU. This work has been carried out within the framework of the EUROfusion Consortium. The views and opinions expressed herein do not necessarily reflect those of the European Commission.

# REFERENCES

[1] M. Lehnen et al, J. Nuclear Materials 463 39 (2015)

[2] J. W. Connor and R. J. Hastie, Nuclear Fusion, 15 415 (1975).

[3] C. Paz-Soldan et al, A novel path to runaway electron mitigation via current-driven kink instability, submitted to IAEA 2020.

[4] D. Carnevale et al., Plasma Phys. Control. Fusion 61 014036 (2019).

[5] Esposito B. et al., PPCF, vol. 59, ISSN: 0741-3335 (2016).

[6] J. R. Martín-Solís, J. D. Alvarez, and R. Sánchez, Physics of Plasmas 5, 2370 (1998);

[7] F. Causa et al, Review of Scientific Instruments 90, 073501 (2019).

[8] M. Gobbin et al, PPCF 60 1 (2017).

[9] G. Papp et al, RE generation and mitigation on the European medium sized tokamaks ASDEX Upgrade and TCV, 26th IAEA (2016).

[10] F. Causa et al., Nuclear Fusion 59 4 (2019).

[11] R. W. Harvey, V. S. Chan, S. C. Chiu, T. E. Evans, M. N. Rosenbluth and D. G. Whyte, Phys. Plasmas 7, 4590 (2000).

[12] P. Helander, H. Smith, T. Fulop and L.-G. Eriksson, Phys. Plasmas 11, 5704 (2004).

[13] C. Paz-Soldan et al, Nucl. Fusion **60** 056020 (2020)

[14] E. Giovannozzi et al, Nuclear Fusion **52** 4535 (2005)

[15] K. Gál *et al, Plasma Phys. Control. Fusion* **50** 055006 (2008)

[16] S. Sridhar *et al., Nuclear Fusion* **60** 096010 (2020)

[17] O. Ficker *et al., Nuclear Fusion* **59** 096036 (2019)

[18] Esposito B et al., Dynamics of high energy runaway electrons in the Frascati Tokamak Upgrade Phys. Plasmas 10 2350 (2003)

[19] Z. Popovic et al., Physics of Plasmas **23**, 122501 (2016)

[20] Causa F, Buratti P, FTU Team, Analysis of runaway electron expulsion during tokamak instabilities detected by a single-channel Cherenkov probe in FTU. Nucl Fusion 59:046013 (2019)

[21] V.V. Parail and O. P. Pogutse, Nuclear Fusion 18(3):303 (1978)

[22] C. Liu et al., Phys. Rev. Lett. 120 265001 (2018)

[23] R. Sweeney et al., J. Plasma Phys., vol. 86, 865860507 (2020)

[24] E.D. Fredrickson et al., Nuclear Fusion vol 55, 013006 (2015) C

[25] D.J. Campbell, A. Eberhagen and S.E. Kissel, Nuclear Fusion 24 297 (1984)

[26] H. Knoepfel and S.J. Zweben, Physical Review Letter, vol 35, n. 20 (1975)

[27] E.M. Hollmann et al, Nuclear Fusion 53 083004 (2013)

[28] L. Carbajal et al., Physics of Plasmas 24, 042512 (2017)

[29] X. Zhu *et al, Nucl. Fusion* **60** 084002 (2020)

[30] S. Cesaroni et al., Fusion Engineering and Design (166) 112323 (2021)

[31] C. Liu et al., PHYSICAL REVIEW LETTERS 120, 265001 (2018)